\documentclass{ws-ijmpd}

\begin{document}

\markboth{G. F. Marranghello and J. A. de Freitas Pacheco}
{Are Neutron-Rich Elements Produced in the Collapse of Strange Dwarfs ?}

\catchline{}{}{}{}{}

\title{Are Neutron-Rich Elements Produced in the Collapse of Strange Dwarfs ?}

\author{G. F. Marranghello and J. A. de Freitas Pacheco}
\address{Observatoire de la C\^ote d'Azur, Bd. de l'Observatoire, B.P.4229,
  F-06304, Nice Cedex 4, France}
\address{marrangh@obs-nice.fr \\ pacheco@obs-nice.fr}  

\maketitle

\begin{history}
\received{(DAY MONTH YEAR)}
\revised{(DAY MONTH YEAR)}
\end{history}

\begin{abstract}
The structure of strange dwarfs and that of hybrid stars with same baryonic
number is compared. There is a critical mass ($M\approx 0.24M_\odot$) in the
strange dwarf branch, below which configurations with the same baryonic
number in the hybrid star branch are more stable. If a transition occurs
between both branches, the collapse releases an energy of about 
of $3~\times 10^{50}$ erg, 
mostly under the form of neutrinos resulting from the conversion of hadronic
matter onto strange quark matter. Only a fraction ($\sim$ 4\%) is 
required to expel the outer neutron-rich layers. These events may
contribute significantly to the chemical yield of nuclides with $A\ge$ 80
in the Galaxy, if their frequency is of about one per 1500 years. 
\end{abstract}


\section{Introduction}

It has been suggested already in the 70's \cite{bc76,cn77} that, because of
the extreme densities reached in the core of neutron stars (NS), hadrons can melt,
creating a deconfined state dubbed the ``quark-gluon plasma". 
Stars with a deconfined core surrounded by hadronic
matter are called {\it hybrid} stars (HS), whereas objects constituted by
absolutely stable strange quark matter are christened ``strange stars" (SS)
\cite{w84,olinto}. 

A pure SS is expected to have a sharp edge with a typical scale defined by the
range of the strong interaction. It was pointed by Ref.~\refcite{olinto} that surface 
electrons may
neutralize the positive charge of strange quark matter, generating a high voltage dipole
layer with an extension of several hundred fermis. As a consequence, SS are 
able to support a crust of
nuclear  material, since the separation gap created by the strong electric field
prevents the conversion of nuclear matter into quark matter. The maximum density of the 
nuclear crust is essentially 
limited by the onset of the neutron drip ($\rho_d \approx 4\times 10^{11}\,
gcm^{-3}$), since above such a value free neutrons fall into the core 
and are converted 
into quark matter. Stars with a strange matter core and an outer layer 
of nuclear material, with 
dimensions typically of white dwarfs, are dubbed "strange dwarfs" (SD)
\cite{gw92,vgs}.  The stability of these objects was examined 
in Ref.~\refcite{Glendenning}, where the pulsation
frequencies for the two lowest modes $n = 0,1$, as a function of the central density,
were studied. More recently, the formation of deconfined cores has been considered in
different astrophysical scenarios. The hadron-quark phase transition induces a
mini-collapse of the NS and the subsequent core bounce was already 
invoked as a possible model
for gamma-ray bursts (GRB). A detailed analysis of the bounce energetics
\cite{fw98} has shown that the relativistic ($\Gamma > 40$) fraction of the 
ejectum is less than
$10^{46}$ erg, insufficient to explain GRBs. However, as the authors have
emphasized, these events could be a significant source of r-process elements. In 
Ref~\refcite{odd02} a different evolutionary path was considered. The point of depart 
is a HS with a
deconfined core constituted only by {\it u, d} quarks. Then, such a core 
shrinks into a more
stable and compact {\it u, d, s} configuration in a timescale shorter than that of
the overlying hadronic material, originating 
a ``quark nova" \cite{odd02,keranen}. The total
energy released in the process may reach values as high as $10^{53}$ erg 
and $\sim 10^{-2}\,
M_{\odot}$ of neutron-rich material may be ejected in the explosion.

The aforementioned events fail in to explain the GRB phenomenology  but could
shed some light on the provenance of elements heavier than those of the iron-peak.
High densities and temperatures required to produce these elements are usually 
found in the
neutrino-driven wind of type II supernovae \cite{Qian}, although the fine-tuning of 
the wind
parameters necessary to explain the observed abundance pattern is still an
unsolved issue \cite{frt99,tbm01}.

In the present work an alternative possibility is explored by considering a
binary system in which one of the components is a "strange dwarf". If this star accretes
mass what will be its new equilibrium state ? The present investigation indicates that
the star can either evolve in SD branch by
increasing its radius or make a jump to the HS (or SS) branch by undergoing 
a collapse in which
the strange core mass increases at the expense of the hadronic layer. We argue that there 
is a critical mass ($\sim 0.24 M_{\odot}$) below which a jump to the HS branch is 
energetically more favorable. In this case, the released energy emitted mostly
under the form of neutrinos, is 
enough to eject a substantial fraction (or almost completely) of the outer neutron rich 
layers, whose masses are typically of the order of $(2-5)\times 10^{-4}\,M_{\odot}$. 
 This paper is organized as follows:  in Section II, strange dwarf models and 
energetics are presented, in 
Section III, the ejection of the envelope and  abundances are discussed 
and, finally, in Section IV the main conclusions are given.

\section{Strange dwarf models}

A sequence of equilibrium (non-rotating and non-magnetic) models were
calculated by solving numerically
the Tolman-Oppenheimer-Volkoff equations\cite{TOV1,TOV2} (G = c = 1), e.g.,

\begin{equation}
\frac{dp}{dr}=-\frac{[p(r)+\epsilon(r)][m(r)+4\pi r^3p(r)]}{r(r-2m(r))}
\end{equation}

and

\begin{equation}
m(r)=4\pi\int_0^r\epsilon(r)r^2dr \, \, ,
\end{equation}

The deconfined core is described by the well known MIT bag model \cite{MIT}, from which 
one obtains respectively for the pressure and energy density

\begin{equation}
p=-B+\frac{1}{4\pi^2}\sum_f[\mu_fk_f(\mu_f^2-\frac{5}{2}m_f^2)+\frac{3}{2}m_f^4
ln(\frac{\mu_f+k_f}{m_f})]
\end{equation}

and 

\begin{equation}
\epsilon=B+\frac{3}{4\pi^2}\sum_f[\mu_fk_f(\mu_f^2-\frac{1}{2}m_f^2) 
-\frac{1}{2}m_f^4 ln(\frac{\mu_f+k_f}{m_f})]\, \, ,
\end{equation}
where $B$ is the bag constant, here taken to be equal to 60 $MeVfm^{-3}$, $k_f$ is the Fermi
momentum of particles of mass $m_f$ and $\mu_f = \sqrt{k_f^2 + m_f^2}$. The sum is
performed over the flavors {\it f = u, d, s}, whose masses were taken respectively equal
to $m_u$ = 5 MeV, $m_d$ = 7 MeV and $m_s$ = 150 MeV. 

\begin{figure}[th]
\centerline{\psfig{file=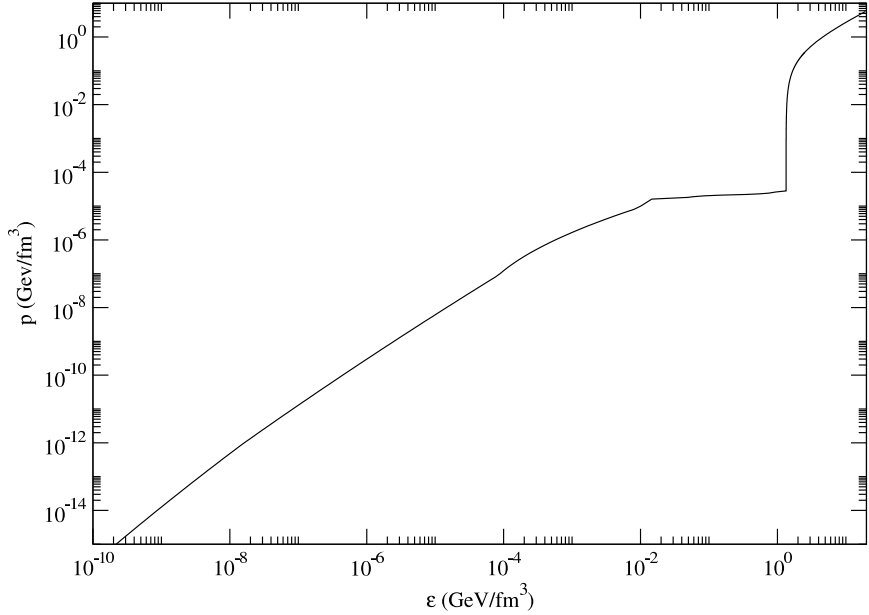,width=10cm,angle=-90}}
\vspace*{8pt}
\caption{The adopted equation of state describing the deconfined core
and the hadronic crust. Both regions are connected at the neutron drip density.
\label{eos}}
\end{figure}

The hadronic layer begins when the pressure at the core radius reach the value corresponding
to the density $\rho_d$ of the neutron drip. The equation of state used for this region
is that calculated in Ref.~\refcite{BPS}. Notice that the bottom of hadronic layer does not represent a 
true phase transition since Gibbs criteria are not satisfied. The strange matter core may absorb 
hadrons of the upper layers if they get in contact, which is precluded by the strong electric field, as
already mentioned. The overall equation of state is shown in Fig \ref{eos}.
At this point, it is important to emphasize the following point. The transition between the 
two phases (deconfined and hadronic) occurs when the Gibbs conditions (equality between the 
chemical potential and pressure of both phases) are satisfied, case of a first order phase 
transition. A mixed phase has also been proposed \cite{Glend92} but the allowance for the
local surface and Coulomb energies may render this possibility energetically less favorable
\cite{heietal}. In the literature, {\it hybrid stars} are those with a deconfined core whose
transition to the hadronic crust is of first order or have a mixed phase. In the present
context, we also call {\it hybrid stars} compact configurations (radii of few km) with a 
quark core and an outer hadronic layer, separated by a strong electric field, as in the
case of strange dwarfs, but this is clearly an {\it abuse} of language.  



\begin{figure}[th]
\centerline{\psfig{file=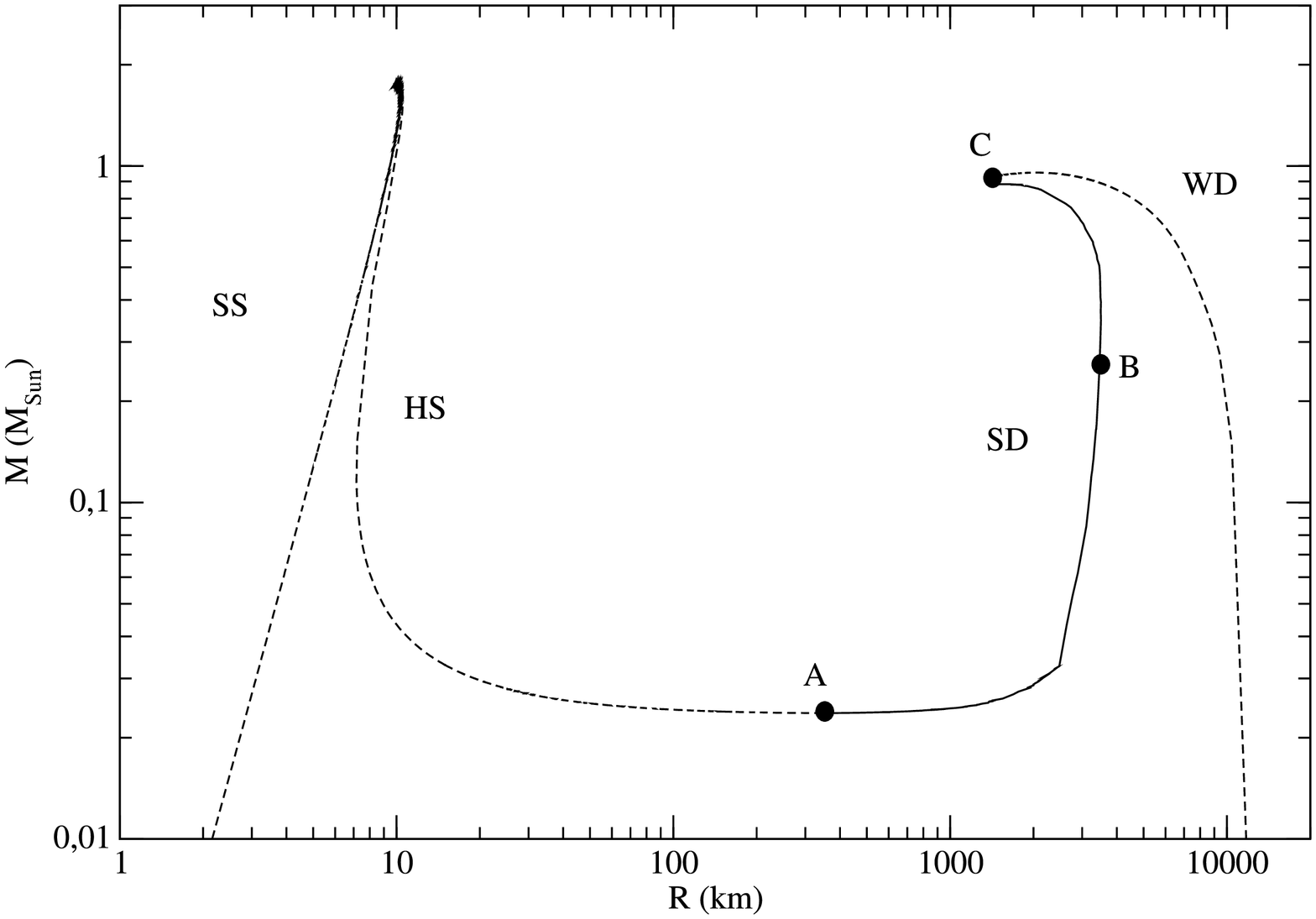,width=10cm,angle=-90}}
\vspace*{8pt}
\caption{Mass-radius diagram for strange dwarfs (solid curve). Other branches
(dashed curves) correspond to white dwarfs (WD), strange stars (SS) and hybrid stars (HS).
\label{f1}}
\end{figure}

The sequence of strange dwarfs and hybrid models was calculated by varying the central energy 
density. In Fig \ref{f1} we show the derived mass-radius (M-R) relation for
our models. The solid curve represents the SD branch, whereas the dashed curves represent respectively the branches
of pure white dwarfs (on the right) and of strange \& hybrid stars (on the
left). 


The point C in the M-R diagram corresponds to the stability edge characterized by a 
mass M=0.80$M_\odot$ and 
a radius R=1397 km. The point A indicates the position of the minimum mass 
of a stable strange dwarf and coincides with the minimum mass model for 
hybrid configurations. It 
corresponds to a mass of 0.0237 M$_{\odot}$ and a radius of 341.82\, km.
As the central density is increased, the star moves along the 
segment A~$\rightarrow$~B in the M-R diagram.
In this range, strange dwarfs have a gravitational mass slightly 
higher than that of hybrid stars of 
{\it same baryonic number}, allowing the possibility for a transition from the SD branch to
the HS branch. This is not the case for strange dwarfs above point B, corresponding to a mass 
of 0.23 M$_{\odot}$, since their gravitational masses are {\it smaller} than 
those of HS stars having
the {\it same} baryonic number. The position of the fiducial points A, B and C 
in the M-R diagram as well
as the curve AC itself depend on the adopted value for the bag constant. A higher value 
would reduce the mass and radius corresponding to point A and similarly, points B 
and C would be displaced toward
smaller masses. This occurs because the role of the strong forces increase, leading
to more compact core configurations \cite{w84}.

Physical properties of some SD and HS models are given in Table I. Models in both branches 
are characterized by a given baryonic mass, shown in the first column. 
The gravitational mass (in solar unit) and radius (in km) for strange 
dwarfs are given respectively 
in the second and third columns, whereas the same parameters for hybrid 
stars are given in columns four and five.
The last column of Table I, gives the energy difference 
$\Delta E = (M_G^{SD}-M_G^{HS})c^2$ between both branches. It is worth mentioning
that $\Delta E$ is the {\it maximum} amount of energy which could be released in
the process. The variation of the gravitational energy is higher but it covers
essentially the cost of the hadronic matter conversion onto strange quark matter. 
Notice that $\Delta E > 0$ for masses lower than $\sim 0.23 M_{\odot}$ and
$\Delta E < 0$ for masses higher than the considered limit, as mentioned above.
The maximum energy difference occurs
around $\sim 0.15 M_{\odot}$, corresponding to $\Delta E \sim 2.9\times 10^{50}$ erg.
Strange dwarfs above point B in the M-R diagram, if they accrete mass, will evolve along the
segment B~$\rightarrow$~C, decreasing slightly the mass and the radius of the deconfined core, but
increasing slightly the extension of the hadronic layer.
The core properties for the same models are shown in Table II. Inspection of this table
indicates that SD along the segment A~$\rightarrow$~C in the M-R diagram have slightly decreasing
deconfined core masses and radii. On the contrary, in the HS branch, the deconfined core develops
more and more as the stellar mass increases.



\begin{figure}[th]
\centerline{\psfig{file=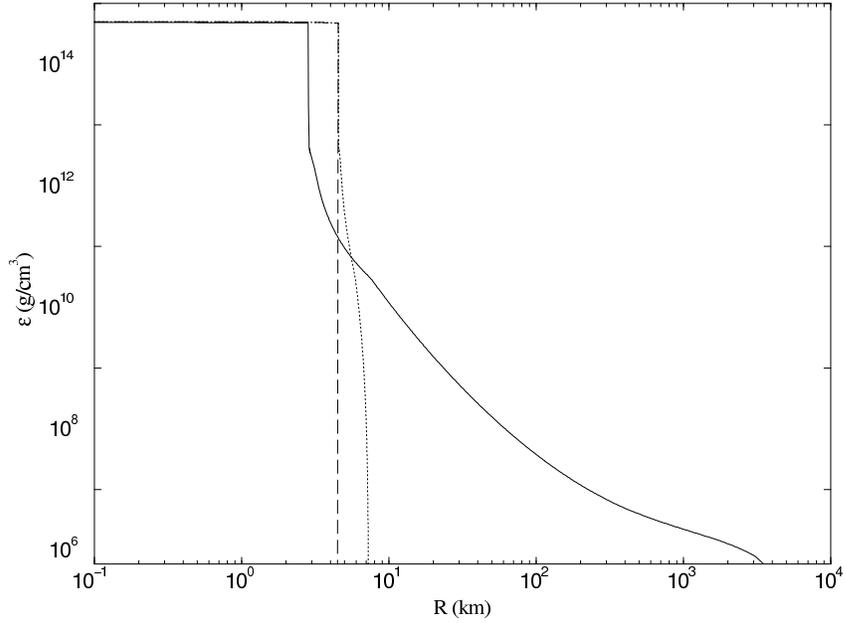,width=10cm,angle=-90}}
\vspace*{8pt}
\caption{Energy density distribution for a strange dwarf (solid line), hybrid
star (dotted line) and a pure strange star (dashed line). All configurations have the same
baryonic mass, $M_B = 0.10696 M_{\odot}$.
\label{f2}}
\end{figure}



\begin{table}[h]
\tbl{Properties of Strange Dwarfs and Hybrid Stars. The last model
corresponds to the minimum mass star and, consequently, has only one possible configuration.\label{t1}}
{\begin{tabular}{|llllll|}\hline
$M_B/M_{\odot}$ & $M_G^{SD}/M_{\odot}$& $R_{SD}$ & $M_G^{HS}/M_{\odot}$ &
$R_{HS}$ & $\Delta E$ \\
\hline
&& (km) && (km) & ($\times 10^{50}$erg) \\ 
\hline
0.40022 & 0.36407 & 3582 & 0.36577 & 8.873 & -30.5\\
0.30543 & 0.27782 & 3547 & 0.27878 & 8.105 & -17.2\\
0.25469 & 0.23165 & 3505 & 0.23179 & 7.721 & -2.50\\
0.20258 & 0.18423 & 3442 & 0.18411 & 7.396 & +2.22\\
0.16808 & 0.15282 & 3357 & 0.15265 & 7.211 & +2.93 \\
0.10696 & 0.09723 & 3119 & 0.09711 & 7.249 & +2.04 \\
0.05185 & 0.04709 & 2556 & 0.04699 & 9.324 & +1.74 \\
0.03626 & 0.03291 & 2309 & 0.03285 & 15.02 & +1.06 \\
0.03000 & 0.02718 & 1756 & 0.02718 & 28.77 & +0.02 \\
0.02613 & 0.02367 & 341.8& 0.02367 & 341.8 &  0 \\
\hline
\end{tabular}}
\end{table}


\begin{table}[h]
\tbl{Core properties of hybrid stars \label{t4}}
{\begin{tabular}{|lllll|}\hline
$M_B/M_\odot$ & $M_{SD}^{core}/M_\odot$ &
$R^{core}_{SD}$ (km) & $M_{HS}^{core}/M_\odot$ &
$R^{core}_{HS}$ (km) \\
\hline
0.40022 & 0.01972 & 2.663 & 0.36560 & 8.288\\
0.30543 & 0.02036 & 2.711 & 0.25832 & 7.135\\
0.25469 & 0.02117 & 2.750 & 0.23167 & 6.471\\
0.20258 & 0.02260 & 2.801 & 0.18401 & 5.923\\
0.16808 & 0.02270 & 2.825 & 0.15257 & 5.202 \\
0.10696 & 0.02277 & 2.842 & 0.09675 & 4.543 \\
0.05185 & 0.02290 & 2.844 & 0.04676 & 3.599 \\
0.03626 & 0.02296 & 2.849 & 0.03187 & 3.169 \\
0.03000 & 0.02298 & 2.851 & 0.02687 & 3.006 \\
0.02613 & 0.02323 & 2.868 & - & - \\
\hline
\end{tabular}}
\end{table}

A comparison between energy density profiles for SD, HS and SS configurations is shown in
Fig.\ref{f2}. All stars have the same baryonic mass ($M_B = 0.10696 M_{\odot}$).
It is also interesting to compare our results with those of Ref~\refcite{vgs}, who have
also performed similar calculations, but with a slight different equation of state for the
quark matter. For their model sequence using the same bag constant and the same
density for the core-envelope transition point, they have
obtained comparable values for the fiducial points defining the SD branch, e.g., 
a mass of 0.017 $M_{\odot}$ and a radius of 450 km for point A and a mass of 0.96 $M_{\odot}$
and a radius of 2400 km for point C. These differences, on the average 20\% in the mass and
30\% in the radius, are probably due to differences in the treatment of the quark matter since
the equation of state for the hadronic crust was taken from the same source. More difficult
to understand are differences in the radii of some configurations which may attain a factor of
three. For instance, their model with a mass of 0.0972 $M_{\odot}$ has a radius
of 10800 km while our calculations for a model of similar mass (0.0973 $M_{\odot}$) give
a radius of only 3130 km.

The analysis of the energy budget of the SD and HS branches suggests that strange dwarfs in
the mass range 0.024 - 0.23 M$_{\odot}$ and in a state of accretion may jump to the HS branch,
releasing an important amount of energy ($\sim 10^{50}$ erg). The following step consists to
study the energetics between the HS and the SS branches. A comparison between models in both
branches indicates that only for masses less than $\sim 0.12 M_{\odot}$ the transition
HS~$\rightarrow$~SS is possible. Characteristics of some computed strange star models are
given in Table III. The first three columns give respectively the baryonic, the gravitational
masses and radii. The last column gives the energy difference $\Delta E = (M_G^{HS}-M_G^{SS})c^2$. 



\begin{table}[h]
\tbl{Strange stars properties. \label{t2}}
{\begin{tabular}{|llll|}\hline
$M_B/M_\odot$ & $M_G^{SS}/M_\odot$ &
$R^{SS}$ (km) & $\Delta E$ ($\times 10^{50}$erg) \\
\hline
0.40022 & 0.39087 & 7.024 & -451\\
0.30543 & 0.27672 & 6.424 & -320\\
0.25469 & 0.23543 & 6.058 & -67.3\\
0.20258 & 0.18521 & 5.608 & -21.4\\
0.16808 & 0.15278 & 5.305 & -0.57\\
0.10696 & 0.09707 & 4.552 & +0.80\\
0.05185 & 0.04699 & 3.609 & +1.84\\
0.03626 & 0.03284 & 3.207 & +1.23\\
0.03000 & 0.02711 & 3.031 & +1.18\\
0.02613 & 0.02367 & 2.876 & +0.15\\
\hline
\end{tabular}}
\end{table}

\section{The neutron-rich envelope}

\subsection{The ejection mechanism}

The astrophysical environment in which slow and rapid neutron capture reactions 
take place is still a
matter of debate. The ejecta of type II supernovae and binary neutron star
mergers are possible sites in which favorable conditions may develop. Difficulties 
with the
electron fraction $Y_e$ in the neutrino-driven ejecta were recently reviewed 
in Ref.~\refcite{pafu00}. 
For instance, a high neutron-to-seed ratio, required for a successful r-process, is 
obtained if the 
leptonic fraction $Y_e$ is small, a condition not generally met in the 
supernova envelope. Non orthodox issues 
based on neutrino oscillations between active and sterile species, able to 
decrease $Y_e$, have been 
explored \cite{pafu00} and here another alternative scenario is examined. 

It is supposed a binary system in which one of the components is a strange dwarf. The
evolutionary path leading to such a configuration does not concern the present work.
As it was shown in the previous section, strange dwarfs with masses in the
range $0.024 < M/M_{\odot} < 0.24$, in a state of accretion, may jump to the HS
branch since this transition is energetically favorable. These considerations 
are based on binding energies calculated for equilibrium configurations
and future dynamical models are necessary to investigate in more detail this possibility.

The jump SD~$\rightarrow$~HS is likely to occur within the free-fall timescale, e.g., 
$t_d \sim 1/\sqrt{G\rho}$, which is of the order of a fraction of millisecond. 
During the transition, hadronic matter is converted onto strange quark
matter. This conversion leads to an important neutrino emission, via weak interaction 
reactions, consisting the bulk of the energy released in the process, which amounts
to about $3~\times 10^{50}$ erg. The typical energy of the emitted neutrino 
pair is about 15-17 MeV,
corresponding approximately to the difference between the energy per particle of the
{\it ud} and the {\it uds} quark plasma. These neutrinos diffuse out the core in a 
timescale of the order of $\sim 0.1 s$ (see, for instance, a discussion in
Ref~\refcite{keranen}) through the remaining hadronic layers, placed above 
the high voltage gap, providing a mechanism able to eject the outer parts of the envelope. 
Masses of the hadronic crust are around $2~\times 10^{-4} M_{\odot}$ and their ejection requires
a minimum energy of about $8~\times 10^{48}$ erg, corresponding to about 4\% of the available
energy.

Neutrinos interact with the crust material through different processes: scattering 
by electrons and nucleons and capture by nucleons. Cross sections for these different
interactions can be found, for instance, in Ref \refcite{Ba89}. The dominant
process in the crust is by far the neutrino-nucleon scattering, whose cross section is
\begin{equation}
\sigma_{\nu-n} = 4~\times 10^{-43}N^2(\frac{E_{\nu}}{10~MeV})^2 \,\,\, cm^2
\end{equation}
where $E_{\nu}$ is the neutrino energy and $N$ is the number of neutrons in the nucleus.
As we shall see below, nuclei in the crust have typically $N \sim$ 35 and 
A $\sim$ 60. Therefore, the ``optical depth" for neutrinos is 
\begin{equation}
\tau_{\nu} = \int \sigma_{\nu-n}(\frac{\rho}{Am_N})ds = 3.3~\times 10^{-18}\int\rho~ds
\end{equation}

The outer hadronic layers have column densities typically of the order of 
$(1.3-3.0)\times 10^{16} gcm^{-2}$, leading to optical depths of the order of
$\tau_{\nu} \sim$ 0.043-0.10, corresponding to a fraction of scattered neutrinos
of about 4.2-9.5 \%. Thus, the momentum imparted to nuclei is able to 
transfer enough energy to expel the envelope. However, a firm conclusion must 
be based on a detailed analysis
of the momentum transfer by neutrinos, coupled to hydrodynamic calculations.

\subsection{Ejected abundances}

The equation of state and the chemical composition of the external 
hadronic matter for densities below the neutron drip were calculated by
different authors \cite{sal,BPS,chung}. Nuclei present in the hadronic crust
are stabilized against $\beta$-decay by the filled electron Fermi levels, becoming
more and more neutron-rich as the matter density increases. The dominant nuclide
present at a given density is calculated by minimizing the total energy density,
including terms due to the lattice energy of nuclei, the energy of isolated
nuclei and the contribution of degenerate electrons, with respect to the atomic
number Z and the nucleon number A.

For a  given model, once the crust structure is calculated from the equilibrium
equations (see Section 2), the mass under the form of a given nuclide (Z,A) can be
calculated from
\begin{equation}
M = 4\pi\int^{R_2}_{R_1}\rho(r,Z,A)r^2dr
\end{equation}
and the integral limits correspond to the density (or pressure) range where the 
considered nuclide is dominant. These nuclides and their respective
density range were taken from tables given by Refs~\refcite{BPS} and \refcite{chung}.
Both set of computations have used similar mass formulas but slightly different
energy minimization procedures. As a consequence, some differences in the
abundance pattern can be noticed. In particular, $_{26}Fe^{76}$ is the dominant
nuclide at densities $\sim 0.4\rho_d$ according to Ref~\refcite{BPS}, whereas
in the calculations by Ref~\refcite{chung} the dominant nuclide is $_{40}Zr^{122}.$

When the envelope is ejected, the  neutron-rich nuclei are no more stabilized
and decay into more stable configurations. Notice that the cross section ratio
between neutrino capture and scattering is $\sim \sigma_a/\sigma_s \approx 0.008$,
indicating that neutrinos will not affect significantly the original abundance 
pattern. Nuclei stability were investigated using a modified Bethe-Weizsacker 
mass formula given in Ref~\refcite{sa02}, more adequate for neutron-rich nuclei,
and nuclide tables given in Ref~\refcite{awt03}.

The resulting masses in the crust for different nuclides are given in Table IV and
Table V, corresponding to the dominant nuclide data by Ref~\refcite{BPS}
and Ref~\refcite{chung} respectively. For both cases, the envelope
mass is $3.6\times10^{-4}M_\odot$. In the first column are given the nuclides 
present in the crust at high pressures, stabilized against $\beta$-decay by
the presence of the degerate electron sea.
In the second column are given the stable nuclides originated from the decay
of the unstable neutron-rich nuclides. The corresponding masses in the
envelope are given in the third column and abundances by number relative to $_{26}Fe^{56}$ 
are given in the fourth column. The last column gives an indication of the expected
origin of these (stable) nuclides in nature: {\it s-} and/or {\it r-} process and
{\it SE} for stellar evolution processes in general, including explosive nucleosynthesis. 

\begin{table}[h]
\begin{tabular}{|ccccc|}\hline
Initial& Final & $M_{eject}$& X/Fe&origin\\ \hline
$_{26}Fe^{56}$&$_{26}Fe^{56}$ & 58 & 1.000& SE\\
$_{26}Fe^{58}$&$_{26}Fe^{58}$& 3& 0.050& SE\\
$_{28}Ni^{62}$&$_{28}Ni^{62}$ & 96 & 1.495& SE\\
$_{28}Ni^{64}$&$_{28}Ni^{64}$ & 55 & 0.829& SE\\
$_{30}Zn^{80}$&$_{34}Se^{80}$ & 17 & 0.205& (s,r)\\
$_{32}Ge^{82}$&$_{34}Se^{82}$ & 25 & 0.294&  r\\
$_{34}Se^{84}$&$_{36}Kr^{84}$ & 36 & 0.413&(s,r)\\
$_{36}Kr^{118}$&$_{50}Sn^{118}$ & 4 & 0.033& (s,r)\\
$_{38}Sr^{120}$&$_{50}Sn^{120}$ & 4 & 0.032& (s,r)\\
$_{40}Zr^{122}$&$_{50}Sn^{122}$ & 13 & 0.103& r\\
\hline
\end{tabular}
\end{table}

\begin{table}[h]
\begin{tabular}{|ccccc|}\hline
Initial& Final & $M_{eject}$& X/Fe&origin\\ \hline
$_{26}Fe^{56}$&$_{26}Fe^{56}$ & 61 & 1.000& SE\\
$_{28}Ni^{64}$&$_{28}Ni^{64}$ & 151 & 2.166& SE\\
$_{30}Zn^{80}$&$_{34}Se^{80}$ & 17 & 0.195& (s,r)\\
$_{32}Ge^{82}$&$_{34}Se^{82}$ & 25 & 0.280&  r\\
$_{34}Se^{84}$&$_{36}Kr^{84}$ & 36 & 0.393&(s,r)\\
$_{36}Kr^{118}$&$_{50}Sn^{118}$ & 8 & 0.062& (s,r)\\
$_{38}Sr^{120}$&$_{50}Sn^{120}$ & 13 & 0.099& (s,r)\\
$_{40}Zr^{122}$&$_{50}Sn^{122}$ & 10 & 0.075& r\\
\hline
\end{tabular}
\end{table}

Inspection of table IV reveals a peak around nuclides in the mass range 56-64 (Fe-Ni peak)
also found in ejecta of type Ia supernovae \cite{Nomoto}. However, the contribution to
the iron yield in the Galaxy by one of these events is about $10^4$ times less
than a single type Ia supernova. Nevertheless, in spite of the the small mass of the
ejected envelope, these events could contribute to the chemical yields of some
nuclides like Se, Kr and Sn, which are usually supposed to be originated from
{\it s} and {\it r} processes. Here their origin is completely diverse, since
they are the result of the decay of neutron-rich nuclides stabilized by
a degenerate electron sea present in the hybrid star. 

The required frequency of these events, in order that they could contribute
significantly to the chemical yield of the Galaxy, can be estimated by
using the procedure by Ref~\refcite{pcib92}. Assuming that all iron in the 
Galaxy was produced essentially by type Ia supernovae and adopting for
Se, Kr and Sn, nuclides which are here supposed  to be producd by the collapse
of a SD, the present abundances given by Ref~\refcite{sprocess},  then the 
required frequency of 
these events in the Galaxy is about one each 1500 yr.

\section{Conclusions}

Gravitational masses for a sequence of models in the strange dwarf, hybrid and strange star
branches were computed. Results of these calculations indicate that there is a critical
mass in the strange dwarf branch, $M = 0.24\, M_{\odot}$, below which a configuration
of {\it same} baryonic number in the hybrid branch has a smaller energy, allowing
a transition between both branches.

If a transition occurs, the envelope radius shrinks typically from a dimension of about
$\sim$ 3200 km to about $\sim$ 7 km, with conversion of hadronic matter onto strange
quark matter. In this collapse, the released energy is about $3\times 10^{50}$ erg
carried out essentially by $\nu_e\bar\nu_e$ pairs with energies typically of
the order of 15-17 MeV. This value corresponds approximately to the energy per
particle difference between {\it ud} and {\it uds} quark matter. Our estimates
indicate that neutrino-nucleon scattering can transfer about 4-9 \% of 
the released energy to nucleons, which is enough to expel partially or
completely the hadronic crust, having masses typically of about of
$(2-5)\times 10^{-4}\,M_{\odot}$. 

The ejecta of these events is rich in nuclides of high mass number and could
be the major source for the chemical yields of elements like Se, Kr, Sn, if the
frequency of these events in the Galaxy is about one per 1500 yr.

\section{Acknowledgements}

GFM thanks the Brazilian agency CAPES for the financial support of this project.

\end{document}